\documentclass[runningheads]{llncs}
\usepackage[T1]{fontenc}
\usepackage{graphicx}
\usepackage[]{todonotes}
\begin{document}
\title {Different Facets for Different Experts: A Framework for Streamlining The Integration of Qualitative Insights into ABM Development\thanks{Supported by the Irish Research Council via Grant COALESCE/2021/4}}
%
\author{Vivek Nallur\orcidID{0000-0003-0447-4150} \and
Pedram Aghaei\orcidID{0009-0008-6316-5335} \and
Graham Finlay\orcidID{0000-0002-4798-2393}}
\authorrunning{V. Nallur et al.}
%
\institute{University College Dublin, Dublin, Ireland\\ 
\email{\{vivek.nallur, pedram.aghaei, graham.finlay\}@ucd.ie}}


%
\maketitle              
\begin{abstract}
A key problem in agent-based simulation is that integrating qualitative insights from multiple discipline experts is extremely hard. In most simulations, agent capabilities and corresponding behaviour needs to be programmed into the agent. We report on the architecture of a tool that disconnects the programmed functions of the agent, from the acquisition of capability and displayed behaviour. This allows multiple different domain experts to represent qualitative insights, without the need for code to be changed. It also allows a continuous integration (or even change) of qualitative behaviour processes, as more insights are gained. The consequent behaviour observed in the model is both, more faithful to the expert's insight as well as able to be contrasted against other models representing other insights. 


\end{abstract}
\section{Introduction}
The ideal that agent-based modelling (ABM) strives to achieve, in many cases, is a true representation of the `society-of-agents' under study, so that we may gain insight into (or even generate) surprising interactions, emergent behaviour, and some level of explainability in an otherwise complex scenario. This promise has led ABM to be used in many and varied domains, e.g., GIS and socio-ecological modelling~\cite{heppenstall2011agent}\cite{filatova2013spatial}, migration networks~\cite{smith_modelling_2014}\cite{klabunde_computational_2018}, epidemiological and crisis simulation~\cite{roche_agent-based_2011}\cite{kruelen_how_2022}, computer games~\cite{meyers_computer_2012}, pedestrian dynamics ~\cite{alqurashi_multi-class_2017}\cite{karbovskii_multimodel_2018}, self-adaptive software~\cite{nallur_clonal_2016}\cite{song2015architectural}, modelling emergence\cite{nallur2016algorithm}, emotion modelling~\cite{horned_models_2023}\cite{amigoni_anxiety_2024}.

For any domain that relies on the agents in the ABM to exhibit realistic behaviour for making predictions (or explanations) about the agents themselves (as opposed to, say, flocking models that seek to explain emergent behaviour), the trueness of representation is extremely critical. However, making true representations of human beings requires expertise from multiple fields, cognitive science, sociology, psychology, etc. Unfortunately, agent-based modelling mechanisms are rarely built to accommodate multiple different experts. To add an additional wrinkle, the output of the model sometimes needs to be interpreted or evaluated by a completely different expert. This is especially true in the field of social simulation, where the human needs to be modelled by (say) sociologists, or transportation experts or health experts, while experimental output needs to be ingested by policy experts. However, this process of `bringing together' of multiple research methods (often called "mixed methods") can be an involved endeavour,  with different disciplines contributing different problem perceptions.
The programmer involved in creating the ABM either needs to acquire expertise in multiple domains, or seeking advice from multiple experts during the lifecycle of the ABM development and usage. A common obstacle faced by agent-modellers is that many aspects of the subject being modelled are often described qualitatively. Depending on the discipline, the translation of qualitative concepts into model parameters or rules may be extremely difficult. Even if somehow translated, there could be loss of nuance due to simplifying assumptions that fail to fully capture the richness and complexity of the subject under study. This is disheartening, since capturing these nuances in simulation are one of the key purported advantages of agent-based modelling.

This paper reports on an attempt to allow different experts to influence and use the ABM, without the programmer  intervening. Specifically, we present the architecture of an ABM that looks at economic migrants into Ireland, and allows ethnographers to use their qualitative knowledge in shaping how the agents (migrants) behave in the model, and also allows policy experts to compare and contrast multiple policy interventions side by side, without being aware of the programmer or the ethnographer. A follow-on advantage of this disconnect is that this architecture allows qualitative insights to evolve  as more data or more experts become available. Note, we do not talk about the specific usecase (economic migration), apart from its use as exemplars for various parts of the architecture.

In the sections that follow, we first (very briefly) describe the problem of integrating qualitative insights, describe the architecture, and the process by which different experts can influence the model. Finally, we describe how domain experts can add nuanced behavioural rules for sidestepping or reinforcing behaviour that has been `coded-in' by the programmer.

\section{Integrating Qualitative Information Into ABMs}
There are many problems that render adding qualitative information as a first-class knowledge source in agent-based modelling, difficult if not impossible. Here, we provide a (very) brief listing of these problems. 
\subsection{Common Problems}
\begin{itemize}
    \item \textbf{Translating Qualitative Concepts into Quantifiable Model Parameters and Rules}: Qualitative concepts are often context-dependent and therefore challenging to be summarized in variables. Increasing the number of variables often makes the model confusing, while decreasing the number of variables can make it too simplistic.
    \item \textbf{Loss of Nuance during Translation}: Existing data may not be directly applicable to the specific context of the model; converting concepts into variables may require transformations that result in a loss of information.
    \item \textbf{Subjectivity and Interpretation}: Almost by definition, there are a multiplicity of cognitive biases and social influences that impact agent decision-making. Distilling these into rules requires making a decision on which perspective to prioritize and which ones to ignore. Once codified, these are implicitly present in all simulation runs.
    \item \textbf{Lack of Validation Data}: When dealing with social, psychological, or other processes that are not fully captured by data, it is difficult to calibrate models to ensure that they reflect reality
    \item \textbf{Overfitting Risk}: In the presence of uncertainty and incomplete information, there is a temptation to lean towards completeness in datasets which may induce the risk of overfitting to a specific set of data points.
    \item \textbf{Disconnect from Empirical Evidence}: Qualitative data alone does not provide information about the relative importance, frequency, or impact of different factors. This can make calibration difficult.
    \item \textbf{Moving Target}: Qualitative insights evolve as new information appears and the society-under-study also reacts to the world around it. At best, this may require certain concepts/variables to be updated, and at worst it may require the model to be completely re-built. We focus on this issue in the current paper.
    \item \textbf{Computational Intractability}: Depending on the model, certain aspects of the agent may be critical (e.g., emotion) to a high-fidelity simulation, but these may be computationally intractable to model
\end{itemize}

\subsection{Typical Integration Approaches}
Depending on the problem, the community has evolved several techniques that allow multiple domain experts to help the developer create a richer agent-based model. Some of these techniques (barbarically summarized for conciseness) are:
\begin{itemize}
   \item \textbf{Participatory Modelling}~\cite{voinov2016modelling}: Involving stakeholders, domain experts, and end-users in the model development process to incorporate their knowledge and perspectives. 
Using techniques like focus groups, interviews, and workshops to elicit qualitative data and translate it into model inputs and assumptions. Hopefully, these result in nuance being captured.
   \item \textbf{Qualitative Data Analysis}~\cite{schluter2017framework}\cite{schluter2019toward}: Qualitative data, such as interview transcripts, field notes, and documents, can be analyzed using techniques like thematic analysis or grounded theory to identify key concepts, relationships, and patterns. These insights can then inform the agent-based model development.  This technique attempts to account for subjectivity.
   \item \textbf{Narrative-Based Modelling}~\cite{elsawah2015methodology}: Incorporating qualitative narratives, stories, and case studies into the agent-based model development process.
   \item \textbf{Hybrid Modelling}~\cite{wijermans_combining_2022}: Combining agent-based models with other modeling techniques like system dynamics, discrete event simulation, or equation-based models to leverage the strengths of different modeling paradigms.
Integrating qualitative data and expert knowledge into the design and parameterization of the agent-based model, attempts to address the problem of adding qualitative insights along with the problem of computational tractability.
\end{itemize}

\subsection{Separated Experts}
A common theme to note among the approaches listed earlier, is that they require multiple experts to come together, and achieve a common understanding of the agent-based model being built. This is a good thing, as the involvement of multiple stakeholders and perspectives, not only brings richness to the model but can also increase the belief in the utility (and acceptance) of the ABM. Unfortunately, this idealized confluence of experts does not always occur. This could be due to multiple reasons. It could be infeasible (financially or schedule-wise) to bring stakeholders from different geographical locations to a single place. There may be organizational constraints which preclude such teaming up. Yet another obstacle is temporal distance. As the real world changes with time, the qualitative insights applicable to the model will change as well. An ethnographer may want to include the latest insights into the simulation run. The end-user of the ABM, perhaps a policy-maker, might want to try out different interventions depending on various options available at the time. This \textit{kink-in-time} is a characteristic of ABM creation and usage in many domains. It is this particular problem of \textit{Moving Target}, that we attempt to address via our facet-based architecture and BehaviourFlow.

\section{Architecture}
The architecture is designed in order to allow different roles to participate in the ABM creation, modification and usage at different times. That is, different roles have the ability to influence different aspects of the ABM without \textit{necessarily} needing the ABM developer to give effect to this influence. In the pursuit of this role separation, we propose a facet-framework that enables the dynamic restructuring of distinct aspects of the model. Although, the concepts we outline would work for any agent-based model, the examples we use henceforth in this paper are derived from our own economic migration model. 
\subsection{Facets (Role: ABM Programmer)}
\begin{figure}
    \centering
    \includegraphics[trim={0.0cm 0.5cm 0.0cm 0.5cm},clip,scale=0.3]{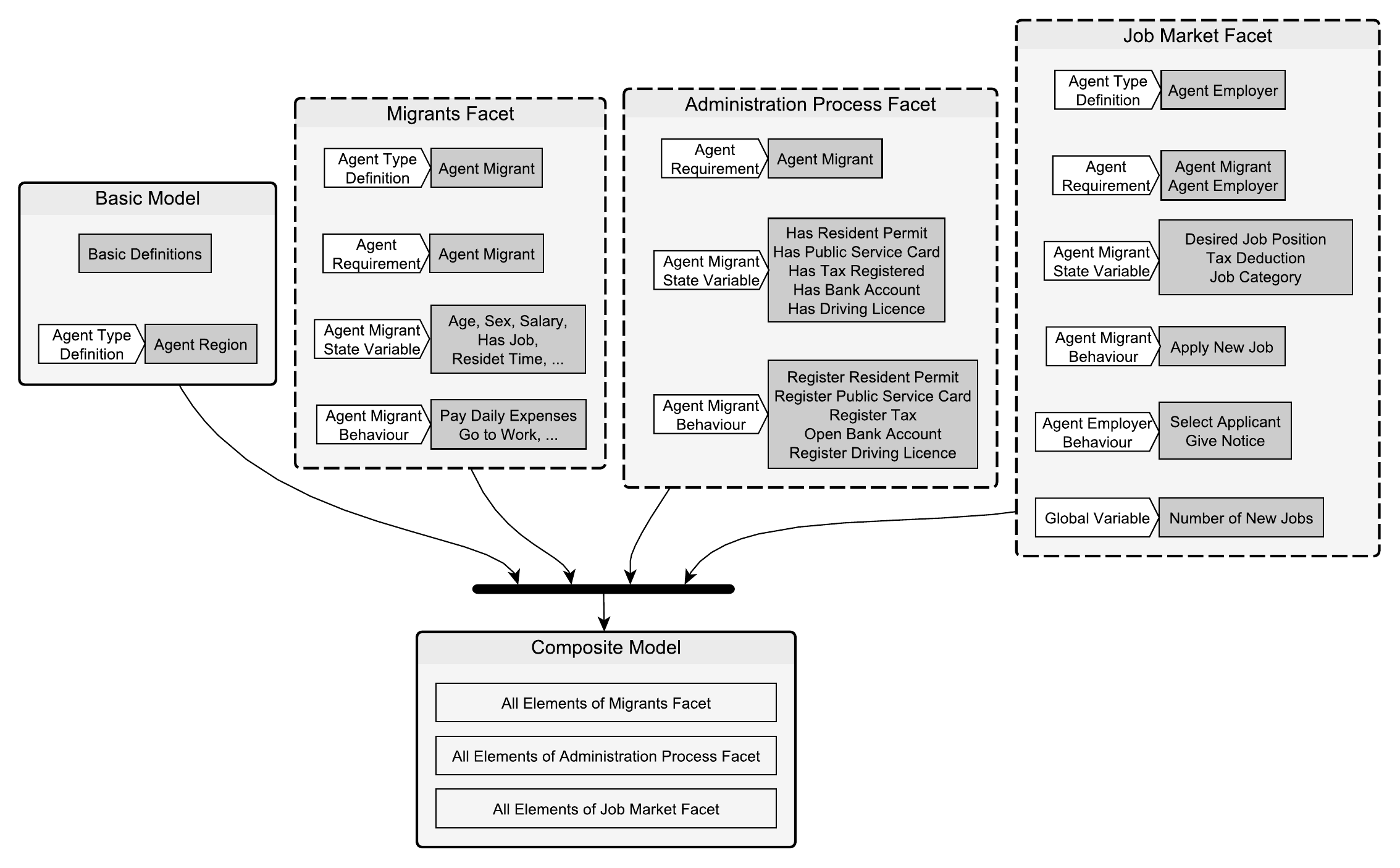}
    \caption{How new facets restructure the model} \label{new_facets}
    \end{figure}

The programmer first creates a Model (This is the way that most ABM programming tools, such as NetLogo or Jason or Mesa, function). This `base model' can be run \textit{as-is}. However, the programmer is able to create facets. A facet introduces a novel set of features or behavioural characteristics to the model. This can be in the form of additional state variables, additional behaviours or even additional agent types. This arrangement of agent capabilities (or decision-making, or communication, or types --- whatever the facet is used for) differs from a typical `layering' mechanism in ABMs. Layers typically add functionality but are all `baked in' during the development process. Facets are available to be included/excluded by a different expert, even after model development is complete, or even deployed. Facets are typically independent domains of action, that may significantly affect agent-behaviour. That is, an expert on the housing market could create a real-estate market facet to reflect their insights of how house availability changes with respect to summer tourist demand, or an ethnographer could use the job-seeking facet to create a functionality specifically for economic migrants from South Asia or Africa. As seen in Figure~\ref{new_facets}, the Base Model (denoted with a continuous line), is augmented with a \texttt{MigrantFacet} (denoted by a dashed line). This adds a new AgentType, additional state variables, as well as behaviour. The \texttt{AdministrationProcessFacet} does not add a new AgentType, but only adds state variables and new behaviour. The \texttt{JobMarketFacet} adds an interaction between two AgentTypes, EmployerAgent and MigrantAgent. These are combined into a \texttt{Composite Model}, which can then be added to a simulation run. Depending on user needs, some facets may not be needed at all for a particular run. The facet-based mechanism allows different combinations to be chosen to compare-and-contrast.

\subsection{BehaviourFlow (Role: Domain Expert)}
\begin{figure}
    \centering
\includegraphics[trim={0.0cm 0.5cm 0.0cm 0.5cm},clip,scale=0.8]{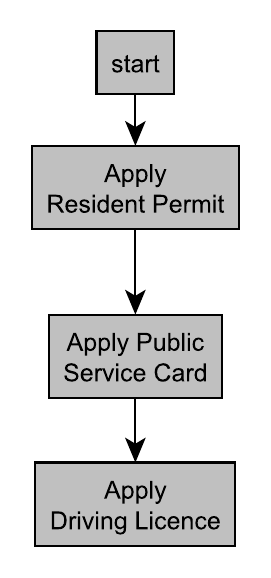}
\caption{A sample BehaviourFlow for a MigrantAgent procuring documents, with the sequence showing the dependency between documents} \label{simple_behaviour_flow}
\end{figure}

The creation of a behaviour flow affects how the agents interact during a simulation run. A BehaviourFlow is how domain experts can insert qualitative insights into agent behaviour. Each AgentType must have a BehaviourFlow. By default, all functions (behaviours) available in an AgentType are available to create a BehaviourFlow. To allow a non-programmer domain expert to design an BehaviourFlow, each function is created as a node in a graph file. This can be manipulated visually, using drag-and-drop editors (such as \texttt{yED}) to create links between the various nodes.
The most basic BehaviourFlow would simply link one node to another in a sequence, as seen in Figure~\ref{simple_behaviour_flow}. This sequential execution would reflect the mechanisms that are typically available in NetLogo or Mesa, where an agent executes behaviour in a step-by-step manner. However, the intention of BehaviourFlow is to allow the domain expert to visually express decision-making by agents (or subsets of agents), taking their particular contexts into account. This decision-making is explained in more detail in Section~\ref{sec_behaviour_flow}

\subsection{Policy (Role: Policy Maker / End-User)}
\begin{figure}
    \centering
\includegraphics[trim={0.0cm 0.5cm 0.0cm 0.5cm},clip,scale=0.45]{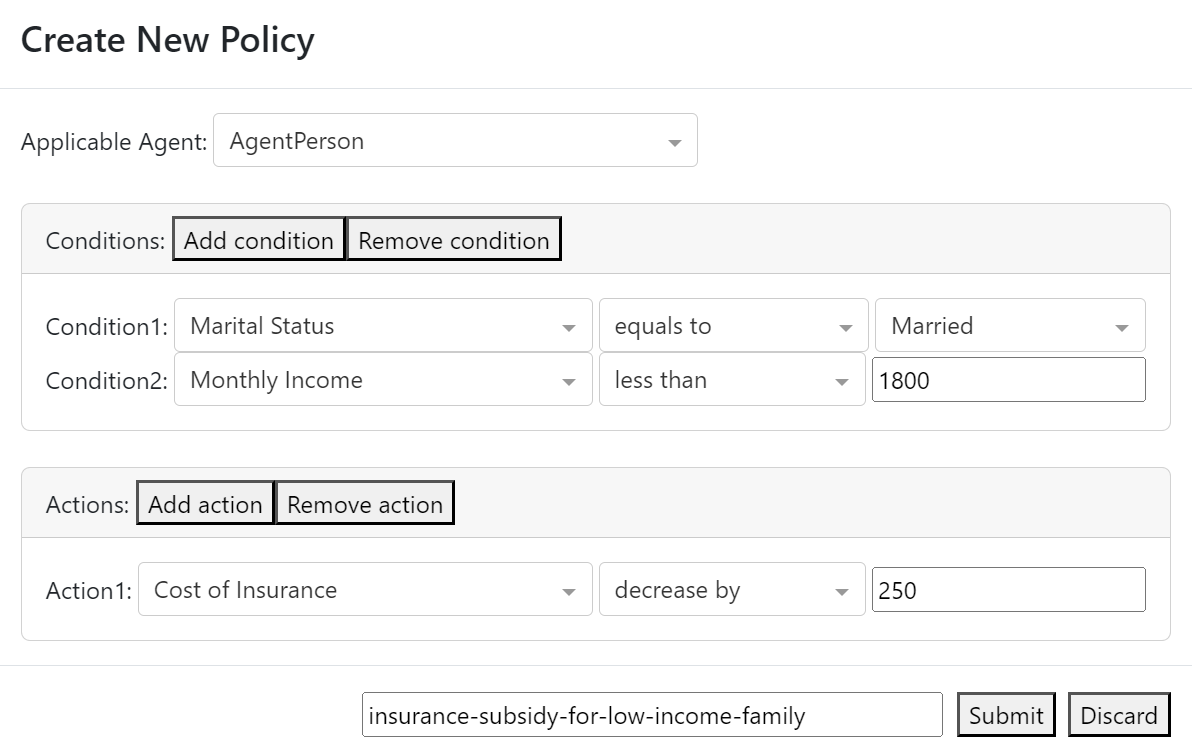}
\caption{Creating new policy by defining conditions and actions} \label{fig_policy_creation}
\end{figure}

During deployment, when the programmer and domain experts are no longer available, it may be desirable for a policy maker to run experiments with different intervention options, \textit{i.e.}, ask counterfactual questions --- if a certain condition were changed, what would happen to the simulation's end result? Unlike typical model parameters, which have to be conceived of in advance, policies are alterations/interventions that can be conceived of by the End-User. For example, in our particular use case, a policy maker could decide to investigate the difference in outcomes if female migrants from a particular region are provided with free public transport \textit{vis-a-vis} migrants from economically depressed regions provided with subsidized medical insurance. As can immediately be surmised, a policy could be highly arbitrary depending on the options available to the policy-maker. Figure~\ref{fig_policy_creation} shows the web-based interface for creating policies. This particular figure shows the creation of a policy that impacts low-income families by providing them with a subsidy on their cost of medical insurance. The Policy Maker is able to create multiple policies and apply them in any combination, as a part of Scenario Design.

\subsection{Scenario Design (Role: End-User)}
\begin{figure}
    \centering
\includegraphics[trim={0.0cm 0.5cm 0.0cm 0.5cm},clip,scale=0.5]{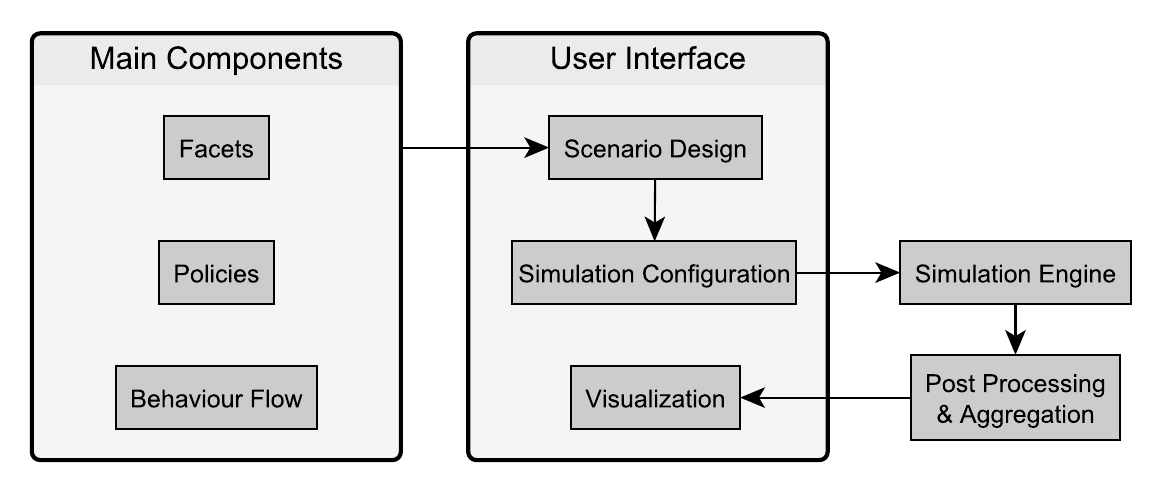}
\caption{Components of the Facet Framework} \label{fig_architecture}
\end{figure}

As seen in Figure~\ref{fig_architecture}, a simulation run consists of a Scenario, that may consist of multiple combinations of Facets, BehaviourFlows, and Policies. These combinations, along with globally affecting variables such as \texttt{number of iterations}, \texttt{data collection intervals}, \texttt{UI parameters}, etc. are grouped into a Scenario. A Scenario can saved, along with its results, so as to enable a comparative re-run. This can be useful if external datasets that feed into a particular AgentType's variables or behaviour (e.g., availability of rentals affecting a HousingAgent) are refreshed, and more current information becomes available.


\subsection*{Current Implementation}
Our proposed framework is built using Python programming language that enables us to exploit the popularity and versatility of Python and its rich packages. The ABM implementation is built on top of the Mesa and Mesa-Geo packages\footnote{https://pypi.org/project/Mesa/}. So, all core components of these packages like the scheduler and data collector are available in this framework. Facets (created by the programmer) are python files, and hence require some programming knowledge. The BehaviourFlow is an XML file that can be visualized using any graph/network diagram editor (such as the freely available \texttt{yED}\footnote{https://www.yworks.com/products/yed/download}). The Scenario Design which consists of policies and global variables, is a JSON file, which can be edited using Web-based UI. The Web-based UI functions as a drop-in replacement for the Mesa visualisation server. This tailored user interface enables policymakers to define different models with various combinations of Facets, Policies and BehaviourFlows and provides visualisation to compare the results, both using charts and choropleth maps. The Web-UI is built using DASH Plotly.

\section{Domain Expert Influencing Agent Behaviour}\label{sec_behaviour_flow}
Now we present a more in-depth discussion of how BehaviourFlow allows an expert to express qualitative insights. Each AgentType has a BehaviourFlow, which decides how instances (agents) of that AgentType behave. At a very rudimentary level, this is similar to the \texttt{go} function in NetLogo or \texttt{step} method that Mesa provides. The difference being that different experts can re-arrange which aspects of the AgentType's behaviour are affected by what conditions. As an example, consider a housing expert's insights into demand for new-build housing by economic migrants with small families (\textit{i.e.}, no children) \textit{vis-a-vis} channeling of financial resources into buying cars by migrants with multiple children. This ethnographic insight emerged, in our case, from various face-to-face surveys combined with statistics of car purchase behaviour in certain areas. However, since these insights change intensity depending on economic region of the country, they cannot be generalized and `coded-in`. Rather, the particular oddities of decision-making are expressed inside a BehaviourFlow. The simple sequential BehaviourFlow with links, as shown in Figure~\ref{simple_behaviour_flow} is quite unrealistic to the manner in which agents actually behave. While the technical aspect of `performing a behaviour' might remain the same, ethnographers' insights into how context is evaluated, whether it rises or drops in priority, what other options are considered, are all important. We realized that even for relatively simpler  flows such as deciding when/whether to apply for a new job, or choosing a particular means of transport to reach their current workplace, there are nuances that continually change as more data is gathered. These nuances in decision-making can be expressed by the domain experts, by visually modifying the (previously simpler) BehaviourFlow. Recall that a node in the BehaviourFlow corresponds to a behaviour that can be executed. A domain expert is able to set BehaviourTriggers on each node. A BehaviourTrigger represents a possibility of the agent enacting the behaviour named by the node. From an implementation perspective, this is a real number between \texttt{zero} and \texttt{one}.  
\paragraph{}
There are three ways to express how this BehaviourTrigger is calculated:
\begin{enumerate}
    \item \textit{Constant}
    \item \textit{Agent's State}
    \item \textit{Arbitrary function}
\end{enumerate}

The constant is used to express behaviours that would (or should) definitely occur for all agents, like updating time. A constant value of $1$ would result in the BehaviourTrigger choosing to execute the behaviour in the node, at every step of the simulation.
\paragraph{}
In some cases, an agent would only execute a behaviour depending on its own internal state. For example, agents that are classified as \texttt{required-to-submit-biometrics} would execute behaviours that result in seeking an appointment, and then attending an appointment. In this situation, the BehaviourTrigger would be set to the Agent's internal state (variable).
\paragraph{}
The most interesting cases occur, where the possibility of execution of behaviour would depend on the evaluation of context. Since context can be arbitrary, the BehaviourTrigger can be set to a function. The function could evaluate multiple variables, examine the agent's own internal state, and then result in a probability, which decides whether a particular agent-behaviour (e.g., \texttt{eating-out} or \texttt{save-money} is executed).
\paragraph{}
While the three mechanisms can express different levels of complexity, sometimes it might be desirable to add multiple criteria. Therefore, a BehaviourTrigger can consist of one or more expressions, involving the above mechanisms. This allows different subsets of agents to select different criteria to decide whether a particular behaviour should be executed or not.  Each expression consists of one or more criterion (\texttt{constant}, \texttt{agent's internal state}, \texttt{arbitrary function}). If all the criteria of an expression are met, the corresponding value (returned by the criteria) is the value of the BehaviourTrigger and determines whether a particular behaviour is executed. To ensure that the simulation runs consistently, a BehaviourTrigger can also set a default value, that is used in case none of the expressions' criteria are met.

\begin{figure}
    \centering
\includegraphics[trim={0.0cm 0.2cm 0.0cm 0.5cm},clip,scale=0.35]{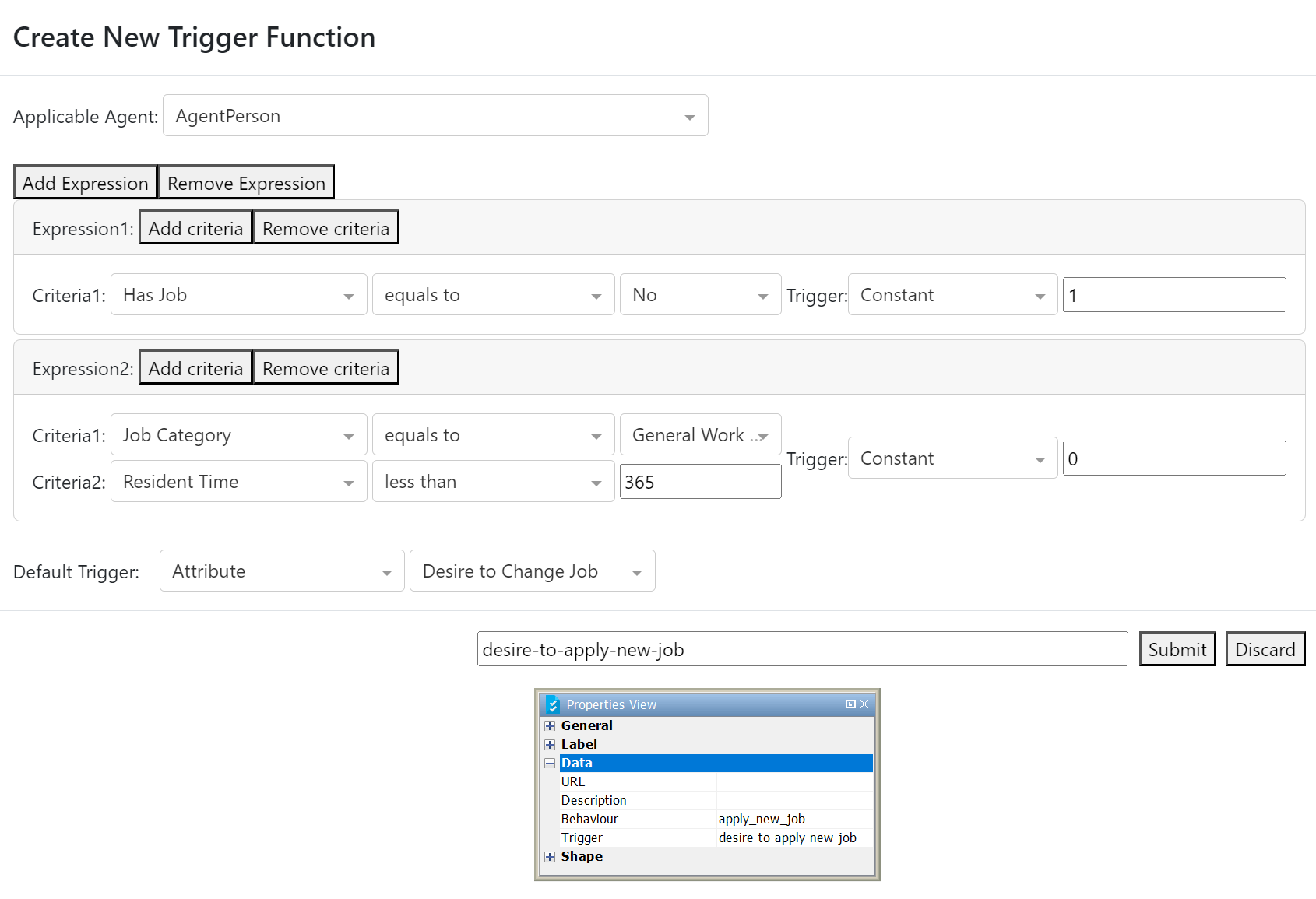}
\caption{A trigger function to define the possibility or desire to apply for a new job} \label{fig_desire_apply_new_job}
\end{figure}

\paragraph{}
Fig.~\ref{fig_desire_apply_new_job} shows an example of a BehaviourTrigger called \texttt{desire-to-apply-new-job} that adjusts the agents' goal-setting behaviour. In this case, \textit{some} agents, depending on their visa work category, are not allowed to seek new employment, unless they have spent a particular amount of time in the country. Whereas others that are not similarly constrained, have their goal-setting determined by whether they already are in possession of a job or not. As can be seen from the figure, the behaviour of goal-setting (and further consequent actions) can be manipulated by quite complex criteria that may be known to sociologists or ethnographers or employment-law specialists, but not necessarily the programmer.

\subsection{Understanding the BehaviourFlow}
For each time step of the model, each agent will begin at the \texttt{start} node of the BehaviourFlow, and proceed through child nodes until it reaches  a node without any children, \textit{i.e.}, a leaf node. Traversing from one node, there could be two conditions. If there is only one child node, the BehaviourTrigger of that node determines whether to execute its behaviour or skip it and proceed to its child. If there are multiple child nodes, the agent chooses one using tournament-selection, \textit{i.e.}, the proportion of the triggers of the child nodes are the weights determining the selection.

This structure allows the implementation of multiple kinds of simulation models. For example, we could create models that contain rules and constraints, entailing certain behaviours as prerequisites for others. For instance, driving a car is conditional upon having a driving license. There may be other prerequisites for applying for a driving license, such as having a resident permit for foreigners, or holding a public services card. 

Another kind of simulation that can be easily modelled is a needs-based model, as described in~\cite{kruelen_how_2022}, that models population (mis)behaviour during a lockdown. The domain expert could design the BehaviourFlow and set the triggers of nodes to reflect the urgency of each behaviour, based on the perceived cognitive needs of the agents. 



\subsection{User Roles and Activities}
\begin{figure}
\includegraphics[trim={0.0cm 0.5cm 0.0cm 0.5cm},clip,scale=0.3]{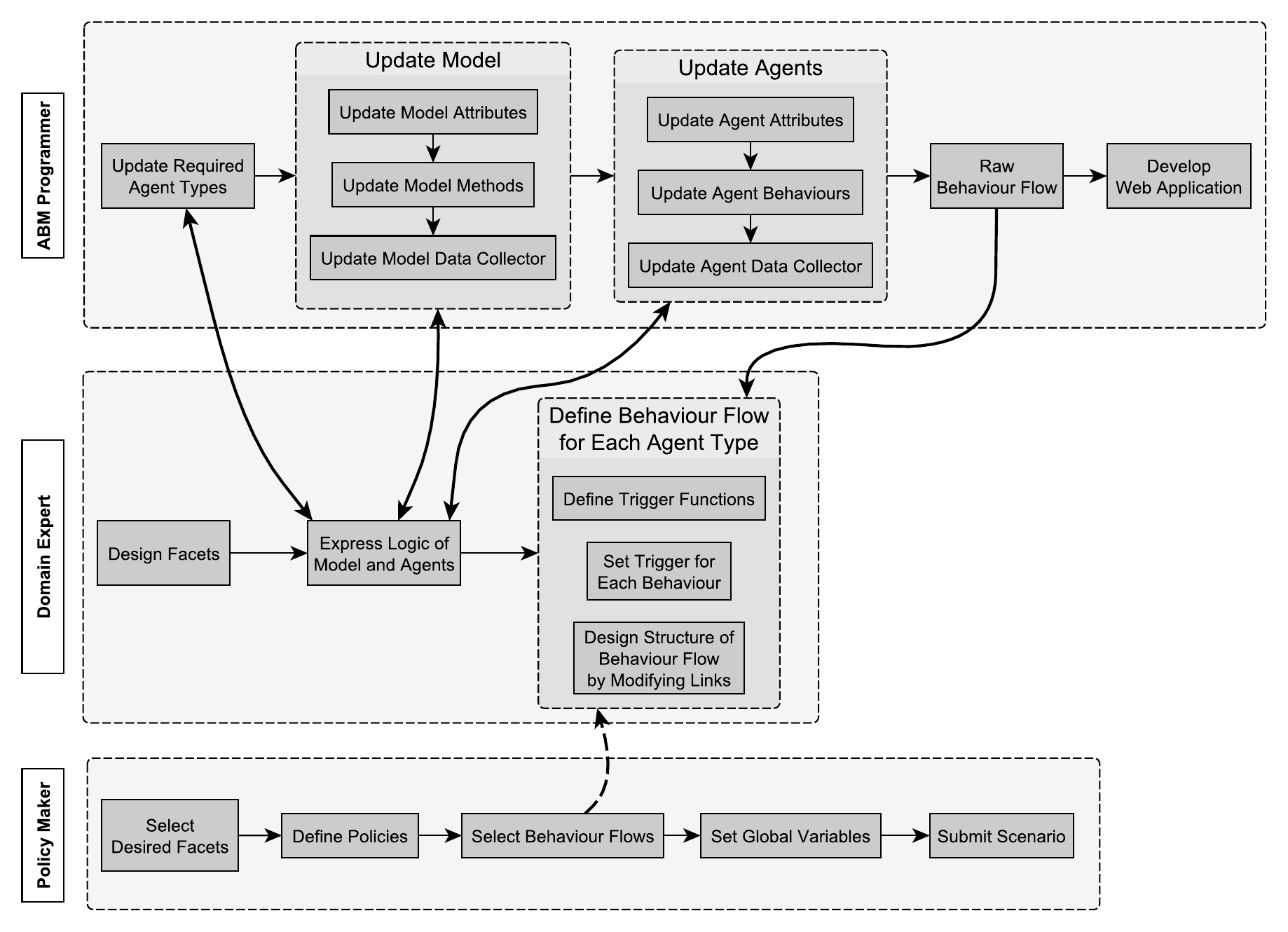}
\caption{A User Workflow} \label{fig_user_workflow}
\end{figure}

Fig.~\ref{fig_user_workflow} shows the workflow that the user would follow to set up a simulation run. Specifically, the figure shows the three roles, the programmer, the domain expert, and the policy maker, being able to modify and influence the simulation. The arrows show the potential for collaborative effort, where different roles can (should) work together to create a richer model. In this particular paper, due to lack of space, we omit discussion of the process of Policy creation and manipulation. However, in principle, the process remains largely the same. The Policy Maker is provided with a Web-based UI that allows them to create policy interventions at the start of a simulation run, and apply them to particular Composite Models. The combination of policy interventions, that affect subsets of agents, along with particular Composite Models are saved as a Scenarios that can be recalled for considered evaluations.

\subsection{Known Weaknesses}
The architecture is flexible enough to accommodate a large range of agent-behaviour modification, targetted interventions to be created without the need for a \textit{`God's eye view}' from the ABM programmer. However, this flexibility creates its own weaknesses. For example, allowing the End-User to create Scenarios requires them to pick facets they would like to include in their simulation. The current implementation, however, cannot detect whether a particular facet has dependencies with other facets. That is, if a complex facet (say, \texttt{HousingFacet}) requires other AgentTypes such as \texttt{SchoolFacet} as well as \texttt{PublicTransportFacet} to function properly, they must be chosen by the End-User. The implementation is unable to warn the End-User of this dependency automatically. 

\paragraph{}Currently, the BehaviourFlow can be edited visually only by external editors. That is, the web application creates a raw BehaviourFlow for each AgentType and saves it as a standard graph format (GraphML - a specialization of XML). The domain expert is required utilize third-party software (such as \texttt{yED} - which is freely available) to drag-and-drop to achieve the correct behaviour required.
\paragraph{}
These are implementation weaknesses, and not architectural. We could envision alternative implementations that are able to account for the weaknesses mentioned above.

\section{Conclusion}
The most important contribution of this work is the separation-in-time that is enabled by the Facet framework. Qualitative experts need not be restricted to explaining finer nuances of their domain to programmers. They are also not restricted to keeping behaviours static, \textit{i.e.}, the BehaviourFlow graphs can be created at any time after the Facets have been created, even during deployment. This flexibility is also available to Policy Makers, who do not need either qualitative experts or programmers to experiment with different options for their simulations. The code for the simulation tool built using this architecture is available, as open-source\footnote{git-url-blinded-for-anonymization}.
\paragraph{}
We wish to be clear that we do not claim that this architecture removes the need for different stakeholders to talk to each other. This is not a methodological substitute for the larger problem of integrating qualitative nuances into an ABM. In fact, we think that multi-stakeholder involvement is an important characteristic and strength of agent-based modelling in social simulations. However, the ability for \textit{some} kinds of decision-making and policy interventions to be modelled, without the need for the programmer being deeply involved, brings benefits. The ability for domain experts to add/modify qualitative insights without depending on a programmer, provides agency, and increases stakeholder involvement. In our particular economic migration modelling usecase, where stakeholders were distributed in space and time, this ability proved to be quite useful.

%
%
\bibliographystyle{splncs04}
\bibliography{references}
\end{document}